# Optimisation and Comprehensive Evaluation of Alternative Energising Paths for Power System Restoration

Shaoyan Li [1,*], Xueping Gu [1], Guangqi Zhou [1], Yang Li [2]

1 School of Electrical and Electronic Engineering, North China Electric Power University, Hebei 071003, China

2 School of Electrical Engineering, Northeast Electric Power University, Jilin 132012, China

*Corresponding author (Shaoyan Li). Email:*shaoyan.li@ncepu.edu.cn

**Abstract:** Power system restoration after a major blackout is a complex process, in which selection of energising paths is a key issue to realize unit and load restoration safely and efficiently. In general, the energising path scheme made beforehand may not be executed successfully due to the possible faults on the related lines under the extreme system condition, so it is necessary to provide alternative path schemes for system restoration. In view of this, the energising path optimisation based on the minimum cost flow model is investigated, then an iterative searching method for alternative path schemes based on mixed integer linear programming is proposed. The iterative method for alternative path schemes could determine more than one scheme with minimal charging reactive power efficiently. In order to make a comprehensive evaluation of the alternative schemes, an evaluation index set is established, and the method based on similarity to ideal grey relational projection is introduced to achieve the final evaluation. The New England 10-unit 39-bus system and the southern Hebei power system of China are employed to demonstrate the effectiveness of the proposed method. The proposed method can provide more efficient and comprehensive decision support for the dispatchers to select reasonable energising paths.

## 1. Introduction

Power system restoration after a major blackout is a vital task for system operation and planning. The whole restoration process can be generally divided into three stages: black-start, network reconfiguration and full load restoration [1-2]. During the network reconfiguration process, energising paths need to be optimised and identified to transfer the cranking power and energise the transmission network [3]. The optimal energising path refers to the optimal transmission line set that connects the black-start unit or the restored zone to one or several target buses. Thus the selection of energising paths is a key issue to realize unit and load restoration safely and efficiently.

Since the 1980s, several approaches about energising path optimisation and determination have been proposed. In [4], a simple and approximate approach for evaluating sustained and transient voltages in energising transmission line is presented. A path selection approach based on power transfer distribution factor (PTDF) and an optimal restoration path algorithm based on "electrical betweenness" are presented in [5] and [6], respectively. But both of the two methods in [5] and [6] focus on restoration of single transmission line without considering the combination effect of the lines in an energising path. Usually, the optimisation of energising path is modelled as a combinatorial optimisation problem solved by graph theory algorithms and the charging reactive power of each line is often assigned as a weighting factor to avoid steady-state overvoltage. The shortest path algorithms, such as Dijkstra algorithm [7-9] and Bellman-Ford algorithm [10], are used to determine the optimal path when there is only one bus to be restored. For the case that all buses need to be connected and restored, the minimum spanning tree algorithm could be used to determine the energising network [10, 11]. Network reconfiguration can be treated as a multi-time-step restoration process [7, 12], in a certain time-step there are usually several plants and substations need to be provided with cranking power via the energising network (path). To determine a minimum weighted subtree connecting the restored zone to several target buses is a typical Steiner tree problem. An approximation method based on Prim algorithm and leaf node pruning strategy is proposed in [13]. In addition to charging reactive power, there are many other factors should be reasonably considered. A new node importance evaluation method is presented based on the concept of regret, and maximisation of the average importance of a path is employed as the objective of finding the optimal restoration path in [14]. Practical factors involving weather condition, equipment reliability, line operation time, and transmission capability are taken into account in [15] to determine the recovery priority of lines. Considering geographic information, an A* algorithm based path search algorithm is proposed in [16].

In recent years, the system restoration optimisation methods based on mathematical programming model show a bright prospect [17-20]. By means of some commercial mathematical programming solvers, like CPLEX [21], the programming model for system restoration can be solved with high computational efficiency. A set of logical constraints between buses and lines for transmission network energisation is identified in [3], the proposed mixed integer linear programming model (MILP) could easily incorporate with other restoration actions such as units restarting [19]. In [20], the proposed transmission line restoration (TLR) model, which is solved by CPLEX solver, is to attain an optimal skeleton-network by identifying the restoration sequence of transmission lines. The connectivity constraint of path restoration optimisation is modelled based on network flow theory in [22]. Besides, other techniques such as, artificial intelligent algorithms [23-24] and expert systems [25] are also tried and implemented to solve the energising path optimisation problem.

The energising path scheme made beforehand may not be executed successfully due to the possible faults on the related lines under the extreme system condition, so it is necessary to provide alternative energising path schemes with priority ordering for system restoration [26]. But most of the current approaches usually only pursue the optimal energising path scheme under certain edge weight settings. The research about the suboptimal alternative path schemes and its comprehensive evaluation is insufficient.

Therefore, a novel method of energising path optimisation is proposed according to the idea of determining the alternative path scheme set first and then sorting the schemes in this paper. The path optimisation based on the minimum cost flow model is investigated, then an iterative searching method for alternative path schemes based on MILP model is proposed. In order to make a comprehensive evaluation of the alternative schemes, an evaluation index set is established. The multiple attribute decision-making method based on similarity to ideal grey relational projection is introduced to achieve the final evaluation and comprehensively optimal scheme. The main advantages of the proposed method compared with the previous methods are that the alternative path schemes with minimal charging reactive power can be determined efficiently by a MILP model and more practical indices are introduced into the scheme evaluation model to achieve the comprehensively optimal energising path.

## 2. Analysis of alternative energising path optimisation problem

### 2.1. Basic process of alternative energising path determination

Network reconfiguration can be treated as a multi-time-stage decision-making process [7, 12], there are usually several important plants and substations which need cranking power for restarting in a certain time-stage. A simple but effective energising path is of great importance for reliably increasing of generation capability quickly. During reconfiguration process, since the restored system is under a light-load condition, excessive reactive power produced by the transmission lines could result in some operational problems, such as steady-state overvoltage and stator end overheating, etc. Thus, minimisation of charging reactive power should be the primary objective of energising path optimisation [27]. Under such circumstances, the path optimisation problem could be modelled as a local minimum spanning tree problem (also called the Steiner tree problem) in graph theory [13, 28, 29].

However, it is not sufficient in path optimisation by only considering minimisation of charging reactive power. Moreover, only one path scheme could not meet the demands of practical systems. Thus, this paper proposes a novel optimisation method of energising paths. First, several path schemes are searched and determined to compose the alternative path set, the required number of schemes are denoted by $M_S$ and the objective is to minimise the charging reactive power of the energising path. Then an evaluation index set is established and a multiple attribute decision-making method is introduced to make a comprehensive evaluation of the path schemes.

Besides, the system conditions after blackout may be different and cannot be known beforehand. Therefore the method must have flexibility and high computational efficiency to adapt to the practical outage conditions. To confirm the flexibility, the proposed method should be able to deal with the following outage condition: there are still several power sources available in the system. For example, the system still has more than one unit in operation due to effective emergency control before blackout.

### 2.2. Mathematical expression of energising path optimisation
#### 2.2.1 Objective:

During network reconfiguration, the restored system is under a light-load condition. Thus the alternative energising path is determined by minimising charging reactive power. If the transmission lines and transformers are equivalent to edges, and the plants and substations are equivalent to buses (nodes), a power system can be treated as a graph $G = (V, E)$ in which $V$ is the set of nodes and $E$ is the set of edges. Then the objective function of energising path optimization can be formulated as:

$$\min \sum_{(i,j) \in E} Q_{Lij} z_{ij} \qquad (1)$$

where the edge $(i, j)$ represents a transmission line or a transformer between bus $i$ and bus $j$; Binary variable $z_{ij}$ indicates the energised status of line $(i,j)$ in proposed energising path optimisation model, where $z_{ij}$ equals 1 if line $(i, j)$ is involved in the energising path, otherwise $z_{ij}=0$; $Q_{Lij}$ is the charging reactive power of transmission line $(i,j)$ and it is used as the edge weight. For a certain line, $Q_{Lij}=B_{ij}V^2$, where $B_{ij}$ is the shunt capacitance of transmission line $(i,j)$, and $V$ is the estimate of voltage level along the energising path [19]. Here, $V$ is set to be 1 p.u.

#### 2.2.2 Constraints:

(1) Radial depth constraint:
To reduce operation risk, the number of stations from the black-start unit or the restored region to any bus to be energised should be limited:

$$\max_{1 \leq i \leq N_D} m_{li} \leq D_{\max} \qquad (2)$$

where $N_D$ is the number of stations or plants to be restored; $m_{li}$ denotes the number of lines in the path from the restored zone to station $i$. For a given path and a given bus to be energized, $m_{li}$ can be obtained by Dijkstra algorithm. $D_{\max}$ is the upper limit of radial depth, which can be adjusted according to the restoration process by the dispatchers. If a bus is too far from the restored zone, it should be restored later.

(2) Reactive power constraint:
The excessive charging reactive power generated by transmission lines may lead to sustained overvoltage since the system is in a light-load state [27]. Therefore the reactive power constraint [30] is defined as follows:

$$\sum_{(i,j)\in E} Q_{Lij} z_{ij} < K_1 \sum_{r=1}^{n_B} Q_{Br,\max} \quad (3)$$

where, $n_B$ is the number of units connected to the restored zone; $Q_{Br,\max}$ represents the maximum reactive power absorbed by unit $r$; $K_1$ is reliability coefficient, its value is less than one. When a certain amount of loads have been pick up with the restoration process proceeding, the value of $K_1$ can be increased appropriately.

(3) Steady state voltage constraints:

The steady state voltage security during the energisation process should be guaranteed:

$$V_{i,\min} \leq V_i \leq V_{i,\max} \quad (4)$$

where $V_i$ is the voltage amplitude of node $i$. This constraint can be checked by calculating AC power flow [13].

## 3. MILP Model for Alternative Path Schemes Determination

As mentioned above, the path optimisation problem can be described as a Steiner tree problem which can be considered as a special case of single commodity uncapacitated fixed charge network flow problem (UFC) in operational research (OR) theory [31]. Therefore, in this section, the standard UFC model is introduced first. Then the problem how to apply UFC model to path optimisation is discussed. Further, the searching and determination of alternative path schemes are studied. Finally, the applicability and generality of the alternative path searching method are analysed.

### 3.1. MILP model for standard UFC problem

UFC is one of a large class of Network Design problems in OR theory. For a given network, there are a set of nodes with flow demand and a set of arcs with fixed and variable cost. UFC problem aims to find arcs and a feasible flow in the resulting network by minimising the total cost. For a graph $G=(V, E)$, the non-negative continuous variable $y_{ij}$ is defined to be the flow on arc $(i, j)$, and the binary variable $z_{ij}$ to be the utilized state where $z_{ij}=1$ if arc $(i, j)$ is used, otherwise $z_{ij}=0$. The MILP model for UFC problem is as follows [31]:

$$\min \sum_{(i,j)\in E} c_{ij} y_{ij} + \sum_{(i,j)\in E} f_{ij} z_{ij} \quad (5)$$

$$\sum_{j\in V_i^-} y_{ji} - \sum_{j\in V_i^+} y_{ij} = b_i, \forall i \in V \quad (6)$$

$$y_{ij} \leq U z_{ij}, \forall (i,j) \in E \quad (7)$$

$$y_{ij} \geq 0, \forall (i,j) \in E \quad (8)$$

$$z_{ij} \in \{0,1\}, \forall (i,j) \in E \quad (9)$$

where, $c_{ij}$ represents the cost of per unit flow and $f_{ij}$ represents fixed cost of arc $(i, j)$; $b_i$ is the flow demand of node $i$; $V_i^+ = \{j \in V : (i,j) \in E\}$, $V_i^- = \{j \in V : (j,i) \in E\}$ represent the set of arcs from node $i$ and the set of arcs to node $i$, respectively; $U$ is a large positive integer, the constraint (7) ensures that $z_{ij}$ takes the value one whenever $y_{ij}$ is non-zero. Note that it suffices to take $U = \sum_{i\in V; b_i>0} b_i$. In addition, define the set $V_S = \{i \in V: b_i < 0\}$ to represent supply nodes, the node in the set is source of flow and its flow demand is negative; the set $V_D = \{i \in V: b_i > 0\}$ represents demand nodes which are the end of flow, the flow demand of node in it is positive; $V_0 = \{i \in V: b_i = 0\}$ is set of transshipment nodes. In a word, the relationship of $y_{ij}$ and $z_{ij}$ can be generalized as follows: $z_{ij}$ takes the value one when $0 < y_{ij} \leq U$; $z_{ij}$ takes the value one or zero when $y_{ij}=0$.

### 3.2. MILP model for the path optimisation problem

To solve the energising path optimisation problem described as Steiner tree problem, some assumptions and settings should be added to the original MILP model for the UFC problem.

1) The flow cost $c_{ij}$ is fixed to zero, and let $f_{ij}$ equal $Q_{Lij}$ (the charging reactive power of the transmission line $i$-$j$). The lines with less charging reactive power have higher priority to be selected to avoid steady-state overvoltage. And binary variable $z_{ij}$ indicates the energised status of the line $(i,j)$ in the proposed model for energising path optimisation.

2) Define $V_P$ as the set of nodes to be connected, set one node in $V_P$ as the only supply node, others are set as demand nodes. The single-source multiple-demand network flow constraints can keep all the nodes in $V_P$ connected;

3) The flow demand of a node in $V_D$ takes value one. In other words, set $b_i=1$ if $i \in V_D$. The flow demand of the only supply node in $V_S$ is set as $b_i = -\sum_{j \in V; b_j > 0} b_j$.

4) The Steiner tree problem is applied in an undirected graph. However, the UFC problem assumes that flow in the arc is non-negative. In order to solve the model easily, the edge in the undirected graph is modelled as two directed arcs.

So the MILP model for Steiner tree problem is as follows:

$$\min \sum_{(i,j)\in E} Q_{Lij} z_{ij} \quad (10)$$

$$\sum_{j\in V_i^-} y_{ji} - \sum_{j\in V_i^+} y_{ij} = b_i, \forall i \in V \quad (11)$$

$$y_{ij} \leq U z_{ij}, \forall (i,j) \in E \quad (12)$$

$$y_{ij} \geq 0, \forall (i,j) \in E \quad (13)$$

$$z_{ij} \in \{0,1\}, \forall (i,j) \in E \quad (14)$$

$$b_i = 1, i \in V_D \quad (15)$$

$$b_i = -\sum_{j\in V_D} b_j, i \in V_S \quad (16)$$

$$b_i = 0, i \in (V - V_D - V_S) \quad (17)$$

### 3.3. Iterative model for alternative path schemes

*3.3.1 Cut for alternative path search:* In order to obtain alternative path schemes, a reasonable idea is to solve the model repeatedly. That is, after an optimal scheme is obtained, it will be removed from the solution space, and the solving process is then repeated to obtain the next optimal solution. The process is repeated until enough schemes are obtained or the model has no more feasible solution. In this way, the alternative path schemes could be found by the iterative searching process. As mentioned above, an energising path can be represented by a group of binary variables $z_{ij}$ corresponding to the energised status of transmission lines. Thus, the removal of the specific energising path from solution space can be realized by adding the following constraint to the original model [31]:

$$\sum_{(i,j)\in E_S}(1-z_{ij}) + \sum_{(i,j)\in(E-E_S)} z_{ij} \geq 1 \quad (18)$$

where, $E_S$ is the set of lines with $z_{ij}=1$ in the energising path scheme determined by (10)-(17); $(E-E_S)$ is the set of lines with $z_{ij}=0$ (lines which are not involved in the energising path with minimal charging reactive power).

Although the constraint (18) can remove the combination of the value of $z_{ij}$ corresponding to specific path scheme from the solution space, the suboptimal scheme from the new search might not meet the practical demand of a reasonable energising path. Its deficiency will be discussed according to the optimal energising path shown in Fig. 1.

It is assumed that the energising path in Fig. 1 is obtained by solving the model (10)-(17). As shown, S and D (including $D_1$, $D_2$ and $D_3$) represent supply node and demand nodes, respectively; the other nodes are transshipment nodes; the number and arrow along the edge show the amount and the direction of flow on the arc indicated by $y_{ij}$. For clarity, the edges that are irrelevant to energising path are not shown here. In other words, only edges with $z_{ij}=1$ and associated nodes are shown in Fig. 1. Through the above analysis, the constraint (18) should be appended to the model of (10)-(17) to obtain the suboptimal energising path. However, three types of invalid suboptimal solutions may be obtained as shown in Fig. 2. In Fig. 2, new edges or nodes are indicated by red bold lines, the edges whose flow is changed are shown by the red number with underline mark.

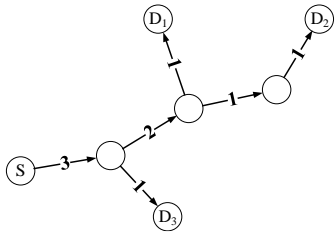

***Fig. 1** Schematic of the optimal energising path*

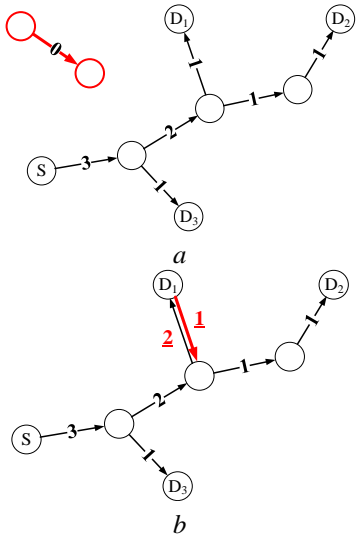

*a*

*b*

*c*

(a) The first type
(b) The second type
(c) The third type

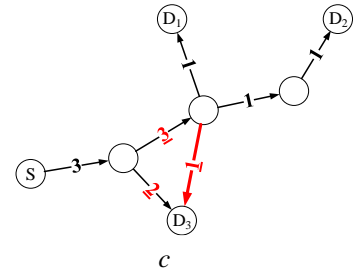

***Fig.2** Three types of invalid suboptimal solutions*

The first type of suboptimal solution is obtained by adding a zero flow edge to the original optimal solution ( the zero flow edge must be the edge with minimal charging reactive power in $(E-E_S)$ ). The second type of suboptimal solution has different network flow with the optimal solution, but the only difference is the addition of the symmetric edge of an edge in the optimal solution. The third type of suboptimal solutions are invalid since they contain a loop structure. By adding constraint (18) to (10)-(17) after the optimal solution is determined, three types of suboptimal solutions shown in Fig.2 will appear. However energising path should be tree type and connected, neither of these three types of suboptimal solutions are valid solutions.

From Fig.1 and Fig.2, an important common feature of all invalid sub-optimal solutions is that they completely include all lines of the optimal solution. However, when a certain energising path has been optimised and determined as the optimal solution, the suboptimal solutions must not include all lines of the optimal solution. Then, a simple but valid linear cut for suboptimal path search is proposed in this paper:

$$\sum_{(i,j)\in E_S}(1-z_{ij}) \geq 1 \quad (19)$$

As mentioned above, the addition of (19) can only exclude a specific energising path corresponding to $E_S$ and $E-E_S$. To remove several energising paths from the solution space, same amount of linear constraints like (19) should be added. The elements of $E_S$ for each iteration change according to its condition.

*3.3.2 Consideration of energised regions:* In the model mentioned above, the function of $y_{ij}$ is only to ensure path connectivity when $c_{ij}$ is fixed to zero, while $z_{ij}$ is the key variable to determine the optimisation objective value. Thus, if $E_{un}$ is defined to be the set of all unenergised lines, the optimal path scheme is actually the combination of lines with $z_{ij}=1$ in $E_{un}$. Considering the effect of restored areas, the objective function can be redefined as:

$$\min \sum_{(i,j)\in E_{un}} Q_{Lij} z_{ij} \quad (20)$$

In summary, the iterative model for alternative path schemes is as follows:

$$\min \sum_{(i,j)\in E_{un}} Q_{Lij} z_{ij} \quad (21)$$

$$\sum_{j\in V_i^-} y_{ji} - \sum_{j\in V_i^+} y_{ij} = b_i, \forall i \in V \quad (22)$$

$$y_{ij} \leq Uz_{ij} \leq Uy_{ij}, \forall (i,j) \in E \quad (23)$$

$$y_{ij} \geq 0, \forall (i,j) \in E \quad (24)$$

$$z_{ij} \in \{0,1\}, \forall (i,j) \in E \quad (25)$$

$$b_i = 1, i \in V_D \quad (26)$$

$$b_i = -\sum_{j \in V_D} b_j, i \in V_S \quad (27)$$

$$b_i = 0, i \in (V - V_D - V_S) \quad (28)$$

$$\sum_{(i,j) \in E_S} (1 - z_{ij}) \geq 1, E_S \in S^m \quad (29)$$

where, $E_S$ is a set of the lines to be restored with $z_{ij}=1$ in the energising path scheme determined by one of the previous $m$ iterations. Obviously, there is $E_S \subseteq E_{un}$. $S^m$ is a set of energising path schemes. In the first iteration there is $S^m = \emptyset$, and in the subsequent iterations there is $S^m = S^{m-1} \cup E_S^{m-1}$, where $E_S^{m-1}$ is the path scheme determined by the ($m$-1)th iteration.

However, since the radial depth and steady state voltage security constraints could not be transformed into linear constraints, these constraints should be considered out of the model presented above. So after each iteration, when the energising path scheme is determined, the scheme is checked to make sure that the reactive power constraint, the radial depth constraint and the steady state voltage constraints are satisfied, then the results can be marked and the next iteration can be performed. If the nodes to be restored are too far from the energised zone or too many nodes are set to be restored, the reactive power constraint may be not satisfied in the first iteration. When the extreme case above occurs, the target nodes should be adjusted. The nodes far away should be restored after the energised system is strong enough.

The whole flow chart of the iterative method is shown as Fig. 3.

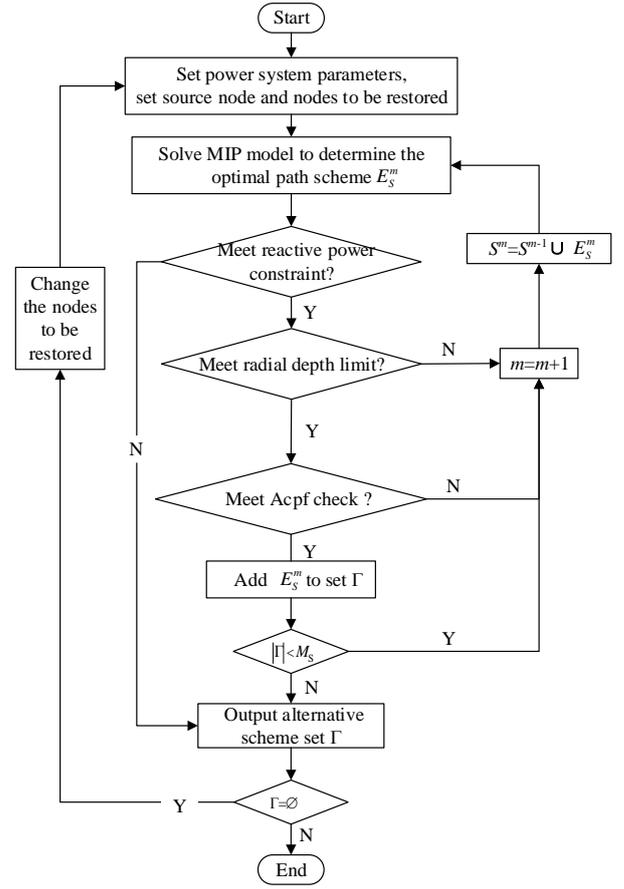

*Fig.3 Flow chart of iterative method for energising paths based on MILP model*

### 3.4. Adaptability analysis of the alternative path searching method

*3.4.1 One target bus case:* When there is only one bus to be energised, the Steiner tree problem is transformed into the shortest path problem. The proposed iterative method based on MILP model can be considered as $K$ shortest path algorithm.

*3.4.2 Multiple energised islands case:* It occurs when the system still has more than one unit in operation due to effective emergency control before the blackout. In this case, in order to determine the optimal path scheme for the unenergised buses, a group of "virtual energised lines" should be added to the original network to connect the energised islands before optimisation. Fig.4 gives a case with three energised islands and its transformation method. Then select an energised node as the only supply node, the multiple islands case could be solved by the iterative model of (21) ~ (29).

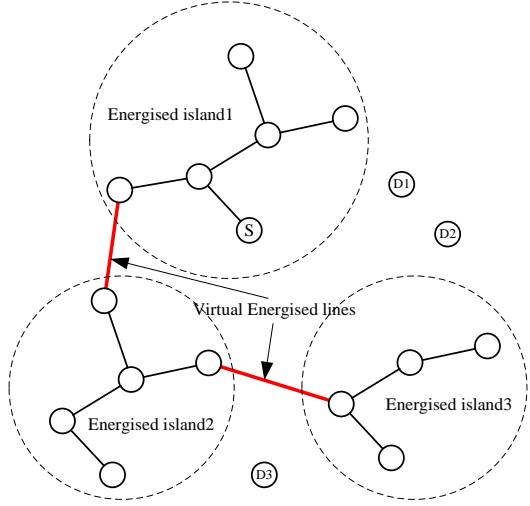

*Fig.4 Transformation of the case of three energised islands*

## 4. Evaluation and Sorting for the Alternative Energising Path Schemes

### 4.1. Evaluation index set of energising path scheme

Although charging reactive power is a primary factor in energising path optimisation, it is necessary to evaluate path schemes comprehensively. And several factors cannot be represented as the weight of the edge. Referring to the optimisation method of black start [33, 34] and network reconfiguration schemes [6, 35, 36], the evaluation index set of the path schemes during the network configuration phase is defined as follows: voltage conversion times $V_1$; switch operation times $V_2$; node importance degree index $V_3$; charging reactive power index $V_4$; radial restoration operation complexity $V_5$. Among the five indices, $V_3$ is a benefit index, and other four are cost indices.

The definitions of these indices are as follows: $V_1$ and $V_2$ are the number of transformers and breakers along the energising path scheme, respectively; $V_3$ is mean value of node importance of all stations on the energising path except restored stations and stations to be restored. $V_3$ can be defined as

$$V_3 = \frac{\sum_{i=1}^{m_n}(I_i + \alpha L_i)}{m_n} \qquad (30)$$

where, $m_n$ is the number of all middle nodes except start node and end nodes; $I_i$ represents the node importance of node $i$ [36]; $L_i$ is the amount of important load on station $i$; $\alpha$ is a positive coefficient determined by expert experience.

$V_4$ is the sum of charging reactive power under rated voltage of all lines belong to the energising path scheme. And it is also the objective value of the model for alternative path optimisation in equation (21). At last, $V_5$ represents radial restoration operation complexity. $V_5$ is defined as:

$$V_5 = \max_{1 \le i \le N_D} m_{li} \qquad (31)$$

where, the meanings of $N_D$ and $m_{li}$ is the same as that in (2); the maximal $m_{li}$ is the radial restoration operation complexity of the scheme. For example, as shown in Fig.1, $N_D$ of energising path scheme is 3, the value of $m_{l1}$ from node S to node $D_1$ is 3, and it is 4 and 2 from node S to node $D_2$ and $D_3$, respectively. Therefore, the value of $V_5$ is 4 in this scheme.

Note that recovery time is also an important index to evaluate the path schemes. But in order to ensure the generality of path evaluation method, it is not considered directly in this paper. Among the evaluation index set, the following three indices are closely related with restoration time and restoration complexity: $V_1$, $V_2$, and $V_5$. If a path scheme has better performance in the three indices above, the recovery time of path will also be short and acceptable relatively.

### 4.2. Comprehensive evaluation of the alternative path schemes based on similarity to ideal grey relation projection

The method to determine attribute weights based on subjective and objective integrated [37] is employed to decide the weights of the five evaluation indices defined above. And then the method of multiple criteria decision-making based on similarity to ideal grey relation projection [38] is employed to evaluate the alternative energising path schemes and determine their order to be used. The decision-making process based on similarity to ideal grey relation projection is as follows.

Supposing $n$ alternative path schemes and $m$ indices construct the original decision-making matrix $A=(a_{ij})_{n \times m}$. In order to unify the different types of attributes into benefit attributes, and normalize the values in the unit interval [0,1], the next formulas are presented for benefit attribute and cost attribute respectively:

for benefit attribute:

$$g_{ij} = \frac{a_{ij} - a_j^{\min}}{a_j^{\max} - a_j^{\min}} \qquad (32)$$

for cost attribute:

$$g_{ij} = \frac{a_j^{\max} - a_{ij}}{a_j^{\max} - a_j^{\min}} \qquad (33)$$

where $a_j^{\min} = \min_{1 \le i \le n} a_{ij}$, $a_j^{\max} = \max_{1 \le i \le n} a_{ij}$, $i=1, \ldots, n, j=1, \ldots, m$; when $a_j^{\min} = a_j^{\max}$, there is $g_{ij}=0$.

In the normalized decision-making matrix $G=(g_{ij})_{n \times m}$, all the attributes have a maximum polarity effect, i.e., the larger the attribute value is, the closer to the goal. The positive ideal scheme $P_0^+$ is defined as a virtual scheme in which each attribute value is the optimal index value of the alternative scheme set, then its index value after normalization is $g_0^+ = \{1, \cdots, 1\}$. Similarly, the normalized attribute values of the negative ideal value scheme $P_0^-$ meets $g_0^- = \{0, \cdots, 0\}$. Then the grey relational coefficient between scheme $P_i$ and the positive (or negative) ideal scheme is as follows:

$$r_{ij}^{+(-)} = \frac{\min_n \min_m \left| g_{0j}^{+(-)} - g_{ij} \right| + \lambda \max_n \max_m \left| g_{0j}^{+(-)} - g_{ij} \right|}{\left| g_{0j}^{+(-)} - g_{ij} \right| + \lambda \max_n \max_m \left| g_{0j}^{+(-)} - g_{ij} \right|} \qquad (34)$$

where $\lambda$ is called distinguishing coefficient, usually, $\lambda = 0.5$; the symbols with superscript +(-) correspond to the positive (negative) ideal scheme. $\left| g_{0j}^{+(-)} - g_{ij} \right|$ is the distance between $j$th attribute value of scheme $P_i$ and the corresponding attribute value of the positive (negative) ideal scheme, and $i=0$ refers to the ideal scheme.

Then the positive ideal grey relational decision-making matrix $R^+$ and the negative ideal grey relational decision-making matrix $R^-$ is defined as follows:

$$R^+ = \begin{vmatrix} r_{01}^+ & r_{02}^+ & \cdots & r_{0m}^+ \\ r_{11}^+ & r_{12}^+ & \cdots & r_{1m}^+ \\ r_{21}^+ & r_{22}^+ & \cdots & r_{2m}^+ \\ \vdots & \vdots & & \vdots \\ r_{n1}^+ & r_{n2}^+ & \cdots & r_{nm}^+ \end{vmatrix}, R^- = \begin{vmatrix} r_{01}^- & r_{02}^- & \cdots & r_{0m}^- \\ r_{11}^- & r_{12}^- & \cdots & r_{1m}^- \\ r_{21}^- & r_{22}^- & \cdots & r_{2m}^- \\ \vdots & \vdots & & \vdots \\ r_{n1}^- & r_{n2}^- & \cdots & r_{nm}^- \end{vmatrix} \quad (35)$$

Further, supposing that the weight set is $w=\{w_1, w_2, ..., w_m\}$, the positive relational projection $Y_i^+$ and negative grey relational projection $Y_i^-$ on the corresponding ideal scheme for scheme $P_i$ are defined as

$$Y_i^{+(-)} = \frac{\vec{P_i} \vec{P_0}^{+(-)}}{\left|\vec{P_0}^{+(-)}\right|} = \frac{\sum_{j=1}^{m} w_j \gamma_{ij}^{+(-)} w_j}{\sqrt{\sum_{j=1}^{m}(w_j)^2}} = \sum_{j=1}^{m} \gamma_{ij}^{+(-)} \frac{w_j^2}{\sqrt{\sum_{j=1}^{m}(w_j)^2}} \quad (36)$$

where $w_j$ is the weight of the $j$th attribute in $w$.

In (36), the bigger $Y_i^+$ is, the closer the scheme $P_i$ is getting to the positive ideal scheme; the bigger $Y_i^-$ is, the closer the scheme $P_i$ is getting to the negative ideal scheme. The ideal scenario is to achieve a bigger $Y_i^+$ and a smaller $Y_i^-$ synchronously, but it is hard to appear. Thus the synthetical grey relational projection $u_i$ is put forward considering both $Y_i^+$ and $Y_i^-$. According to the least square sum criteria, the objective function is defined as follows:

$$\min F(u_i) = \left(u_i \cdot Y_i^+ - Y_i^+\right)^2 + \left[(1-u_i) \cdot Y_i^- - Y_i^-\right]^2 \quad (37)$$

Let $\frac{\partial F(u_i)}{\partial u_i}=0$, the expression of the synthetical grey relational projection $u_i$ will be obtained as follows:

$$u_i = \frac{Y_i^{+2}}{Y_i^{+2} + Y_i^{-2}} \quad (38)$$

Obviously, $u_i$ varies within the range [0, 1], and the scheme with the maximum synthetical grey relation projection coefficient is the final optimal energising path scheme.

## 5. Case Study

### 5.1. Case study I: New England 10-unit 39-bus power system

To explain the features of the proposed method and validate its effectiveness, two scenarios are built based on the New England 10-unit 39-bus system. The first scenario contains one electrical source and multiple buses to be energised, the second scenario contains three energised islands and multiple buses to be energised.

#### 5.1.1 One electrical source scenario

The bus and line numbers of the New England 10-unit 39-bus power system are as shown in Fig. 5. The Unit 33 is assumed to be the black-start unit with an installed capacity of 652 MW and maximum reactive power of 250 MVar, and it has successfully self-restarted. Suppose that the dispatchers decide to establish the energising path for bus 6, bus 15 and bus 17 simultaneously. The parameter $K_1$ is given a value of 0.8, and maximum reactive power $Q_{Br,max}$ absorbed by unit $r$ is $0.3S_{Nr}$, where $S_{Nr}$ is the rated capacity of unit $r$. The maximum reactive power absorbed by black-start unit 33 is 167.59 MVar according to (2). The radial depth constraint of path is set as $D_{max}=8$, and the required number of alternative schemes is $M_S=8$. The iterative model is modelled in GAMS platform [39], and MILP model composed by (21) ~ (29) is solved by CPLEX solver, the AC power flow problem (for the steady-state voltage security check) is solved by IPOPTH solver, the radial depth constraint is modelled in GAMS based on Breadth First Search (BFS). The computations were done on a 3.3GHz Windows 7 PC with 4.00GB of RAM.

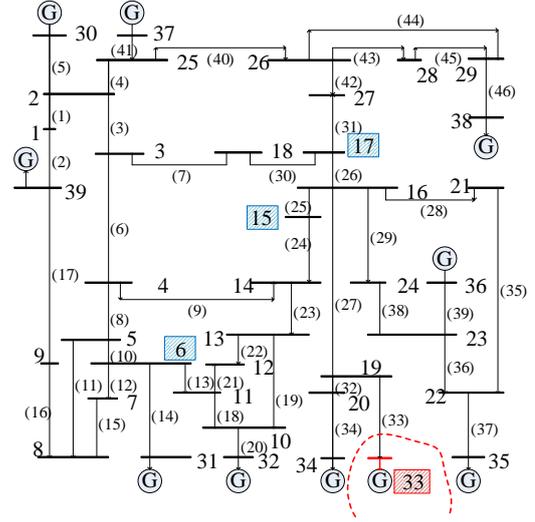

Energised zone    XX Energised buses    XX Target buses

*Fig.5 Network of New England 10-unit 39-bus power system*

By employing the iterative method based on mixed integer linear programming, the alternative path set is obtained in 2s. As shown in Table 1, expect for the scheme 8, the other 7 schemes all satisfy the reactive power constraint. The scheme 8 is invalid because it generate larger charging reactive power than unit 33 can absorb (167.59 MVar).

Meanwhile, when the radial depth constraint, reactive power constraint, and AC power flow constraints are not considered, the iterative model could be used to solve the K minimal Steiner trees problem. In other words, the model could give accurate the first K subtrees with minimal sum of charging reactive power. Thus there are no more schemes that meet the reactive power constraint in this case. When all these constraints are taken into account, only four schemes from No.1 to No.4 in Table 1 are valid with all the constraints satisfied, and they are further illustrated in Fig.6 .

**Table 1** Energising path schemes for one electrical source scenario of New England 10-unit 39-bus power system

| Scheme No. | Lines involved | MVar |
|---|---|---|
| 1 | 13 21 22 23 24 25 26 27 33 | 128.64 |
| 2 | 8 9 10 24 25 26 27 33 | 129.10 |
| 3 | 6 7 8 10 25 26 27 30 33 | 135.39 |
| 4 | 13 18 19 23 24 25 26 27 33 | 143.22 |
| 5(invalid) | 8 9 11 12 15 24 25 26 27 33 | 158.62 |
| 6(invalid) | 6 7 9 13 21 22 23 25 26 27 30 33 | 162.57 |
| 7(invalid) | 6 7 8 11 12 15 25 26 27 30 33 | 164.91 |
| 8(invalid) | 6 7 8 9 24 26 27 30 33 | 168.71 |

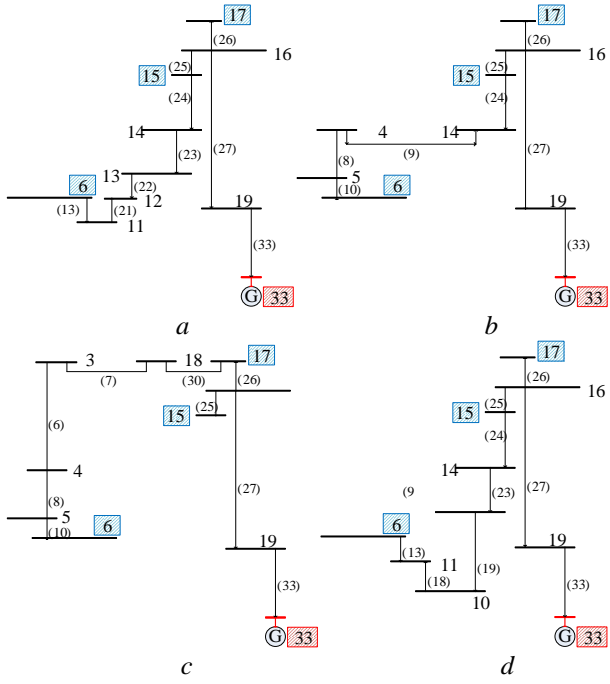

(a) Scheme 1
(b) Scheme 2
(c) Scheme 3
(d) Scheme 4

***Fig.6*** *Schematic diagram of scheme 1-4*

Calculate the comprehensive weights of five indices by the method proposed in Section 4, and the result is $w$= [0.1525, 0.1709, 0.1970, 0.2382, 0.2413]. And then the four valid schemes could be sorted by the method of similarity to ideal grey relation projection. The grey relational projection coefficients and index values of the alternative schemes are shown in Table 2. To illustrate the invalidity of the schemes 5-8, the index values of all the 8 schemes are presented (the over-limit index values are typed in bold and with underline mark).

**Table 2** Index values for one electrical source scenario of the New England 10-unit 39-bus power system case

| Scheme No. | $V_1$ | $V_2$ | $V_3$ | $V_4$(MVar) | $V_5$ | $u_i$ |
|---|---|---|---|---|---|---|
| 1 | 3 | 18 | 0.0063 | 128.64 | 8 | 0.286 |
| 2 | 1 | 16 | 0.0066 | 129.10 | 7 | 0.896 |
| 3 | 1 | 18 | 0.0065 | 135.39 | 8 | 0.358 |
| 4 | 1 | 18 | 0.0064 | 143.22 | 8 | 0.186 |
| 5(invalid) | 1 | 20 | 0.0064 | 158.62 | **<u>9</u>** | - |
| 6(invalid) | 3 | 24 | 0.0063 | 162.57 | **<u>11</u>** | - |
| 7(invalid) | 1 | 22 | 0.0064 | 164.91 | **<u>10</u>** | - |
| 8(invalid) | 1 | 20 | 0.0065 | **<u>168.71</u>** | 8 | - |

According to Table 2, the optimal energising path scheme is scheme 2, and the suboptimal alternative schemes are scheme 3, scheme 1 and scheme 4 by descending order. Though the energising reactive power of scheme 2 is bigger than scheme 1, scheme 2 has a better comprehensive performance. Besides, since scheme 2 has the best performance in $V_1$ (voltage conversion times index), $V_2$ (switch operation times index), and $V_5$ (radial restoration operation complexity index), it can help save valuable time in system restoration. From the results above, the alternative path optimisation method can determine several valid schemes with minimal charging reactive power. However, it is insufficient to evaluate energising path schemes only by charging reactive power, the proposed evaluation indices and method can provide more rational and comprehensive decision-making support during system restoration.

In view of this path optimisation problem for multiple target buses, an approximation method is proposed in [13]: First, construct the minimum spanning tree using the Prim algorithm. Then, remove the edges that don't need to be restored by pruning operation, which guarantees all leaf nodes are the target nodes. In this case, the only scheme obtained by using this approximation method is scheme 3 in Table 1. Obviously, scheme 3 is neither the scheme with minimal energising reactive power nor the scheme with the best comprehensive performance.

In addition, the method of "energising one bus in each step" is developed in [10] to optimise the energising path for multiple target buses. The idea of this method is to divide the path optimisation for multiple nodes to many steps. In each step, only one target node is energised based on the shortest path algorithm. Then, the energising path is identified as a restored state. Based on the modified state, the path to restore next node is determined, and finally the energising path is obtained. It is obvious that the sequence of nodes should be determined in advance. In this case, there are 3 nodes to be restored, which means that six sequences ( $A_3^3=6$ ) of nodes should be considered. Only two different schemes are obtained according to all possible sequences. They are corresponding to the scheme 1 and scheme 3 in Table 1 and Fig. 6. Although scheme 1 has minimum charging reactive power, the comprehensive evaluation value is lower than that of the scheme 2. Therefore, even if all the search sequences of the nodes are traversed by "energising one bus in each step" method, only few schemes can be obtained and the performance of them cannot be guaranteed.

Solution quality and computation time of different methods are compared in Table 3, it can be seen that the proposed method can provide adequate feasible solutions in a short time. The global optimality of schemes obtained by method in [13] and [10] method cannot be assured although the computation time is relatively less. And the method in [13] can only determine one scheme. The method in [10] may give few different schemes according to the specific situation, but the number and optimality of the schemes cannot be guaranteed.

**Table 3** Features of the methods which generate alternative schemes

| Methods | Global optimality | K optimal schemes* | Computation time (s) |
|---|---|---|---|
| Reference [10] | Maybe | No | 0.700 |
| The proposed Method | Yes | Yes | 2.000 |

*The first K schemes with minimal energising reactive power

### 5.1.2 Three energised islands scenario

To further explain the features of the proposed method and validate its effectiveness, a scenario with three energised islands is built based on the New England 10-unit 39-bus system. Before optimisation, line 33-30 and 33-38 are added to the original system as the "virtual energised lines" to connect the islands. At the same time, bus 33 is chosen as the only supply node. The bus and line number of the test system is shown as Fig. 7. Suppose that the dispatchers decide to establish the energising path for bus 6, bus 15 and bus 17 simultaneously. Because several important loads have been picked up, the charging reactive power constraint is not considered in this test case. The radial depth constraint of path is still set as $D_{max}=8$, and the required number of alternative schemes is $M_S=8$.

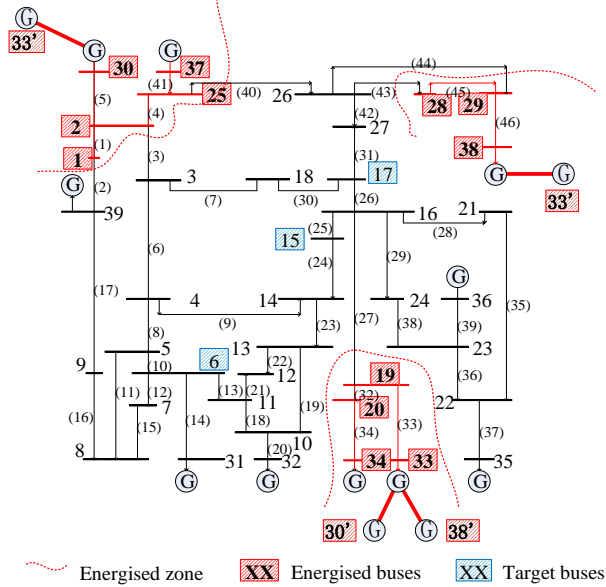

*Fig.7 Network of the test system after equivalent transformation*

The iterative solving process based on the mixed integer linear programming model is carried out, the alternative path set obtained in 1.5s is shown in Table 4 (8 schemes all satisfy radial depth, reactive power, and steady-state voltage constraints).

**Table 4** Energising path schemes for the case of New England 10-unit 39-bus power system

| Scheme No. | Lines involved | MVar |
|---|---|---|
| 1 | 3 6 8 10 25 26 27 | 126.54 |
| 2 | 13 21 22 23 24 25 26 27 | 128.64 |
| 3 | 8 9 10 24 25 26 27 | 129.10 |
| 4 | 3 6 7 8 10 25 26 30 | 130.71 |
| 5 | 6 7 8 10 25 26 27 30 | 135.39 |
| 6 | 13 18 19 23 24 25 26 27 | 143.22 |
| 7 | 3 6 8 9 10 24 25 26 | 146.56 |
| 8 | 3 6 7 8 10 25 27 30 | 147.69 |

Calculate the comprehensive weights of five indices by the method proposed in Section 4, and the result is $w$= [0.1139, 0.1449, 0.1516, 0.2053, 0.3844]. And then the schemes could be sorted by the method of similarity to ideal grey relation projection. The grey relational projection coefficients and index values of the alternative schemes are shown in Table 5.

**Table 5** Index values of path schemes for the New England 10-unit 39-bus power system case

| Scheme No. | $V_1$ | $V_2$ | $V_3$ | $V_4$(MVar) | $V_5$ | $u_i$ |
|---|---|---|---|---|---|---|
| 1 | 0 | 14 | 0.0067 | 126.54 | 4 | 0.900 |
| 2 | 2 | 16 | 0.0064 | 128.64 | 7 | 0.181 |
| 3 | 0 | 14 | 0.0067 | 129.10 | 6 | 0.612 |
| 4 | 0 | 16 | 0.0066 | 130.71 | 5 | 0.639 |
| 5 | 0 | 16 | 0.0066 | 135.39 | 7 | 0.215 |
| 6 | 0 | 16 | 0.0064 | 143.22 | 7 | 0.146 |
| 7 | 0 | 16 | 0.0067 | 146.56 | 6 | 0.364 |
| 8 | 0 | 16 | 0.0066 | 147.69 | 4 | 0.705 |

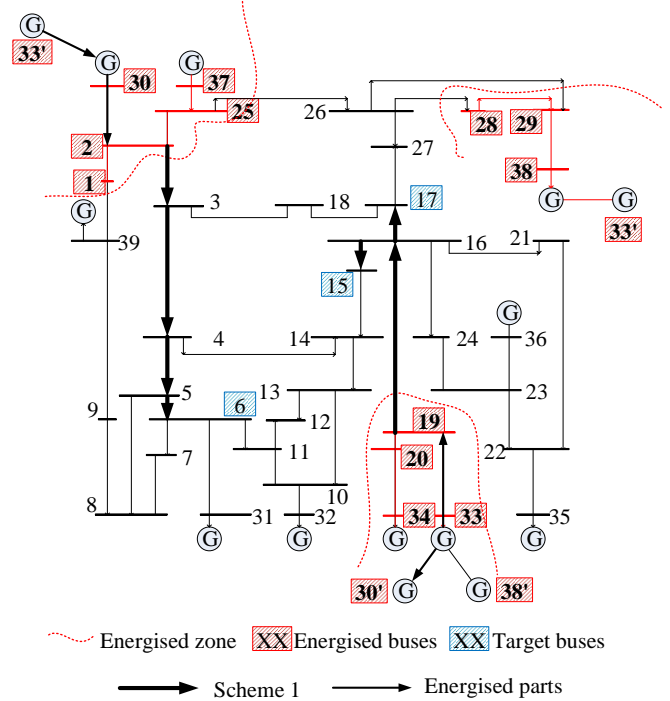

*Fig.8 Schematic diagram of scheme 1*

According to Table 5, the optimal energising path scheme is the scheme 1, and the suboptimal alternative schemes are the scheme 8, 4, 3, 7, 5, 2 and 6 by descending order. The optimal energising path are shown in Fig. 8. Although the charging reactive power is the largest in all schemes, other indices show better performances, especially the radial restoration operation complexity $V_5$. From the results above, the alternative path searching method can solve the multiple energised islands case.

The method based on minimum spanning tree in [13] can not solve multiple energisied islands cases. The method of "energising one bus in each step" proposed in [10] can solve multiple energisied islands cases, but after all the search sequences of the nodes are traversed, only scheme 1 and scheme 4 are obtained. The method proposed in this paper is better than the method in [10] in both the number and performance of alternative schemes.

### 5.2. Case study II: the southern Hebei power system of China

To illustrate the effectiveness of the proposed model in a practical power system, the test is carried out on the southern Hebei power system of China [40]. The system includes 96 stations, 19 generating units, 187 transmission lines with voltages at 220 kV and above. For clarity of description, only 500kV power stations and part of 220kV power stations are shown in Fig. 9. The black-start unit in ZHW has been successfully restarted, the restoration path should be optimised to restore YH, SA, SJZ and XBP simultaneously. The radial depth constraint of path is set as $D_{max}=8$, and the required number of alternative schemes is $M_S=8$. After path searching and evaluation, the optimal scheme and alternative schemes are obtained, and the first 4 alternative schemes with maximum $u_i$ are shown in Fig. 9. Among them, Scheme 1 is the optimal one since which has the maximum $u_i$ and best comprehensive performance.

Besides, the optimal scheme has the minimum switch operation times, and radial restoration operation complexity, thus it can save valuable time to restart units and increase the generation capability of system. The total computational time consuming in this case is about 3s. Thus, alternative schemes re-optimisation can be accomplished in seconds, when the target buses are changed or accidental lines failures occur during the restoration process. The optimisation results demonstrate that the proposed method could solve the energising path optimisation problem for complex outage cases and practical power systems.

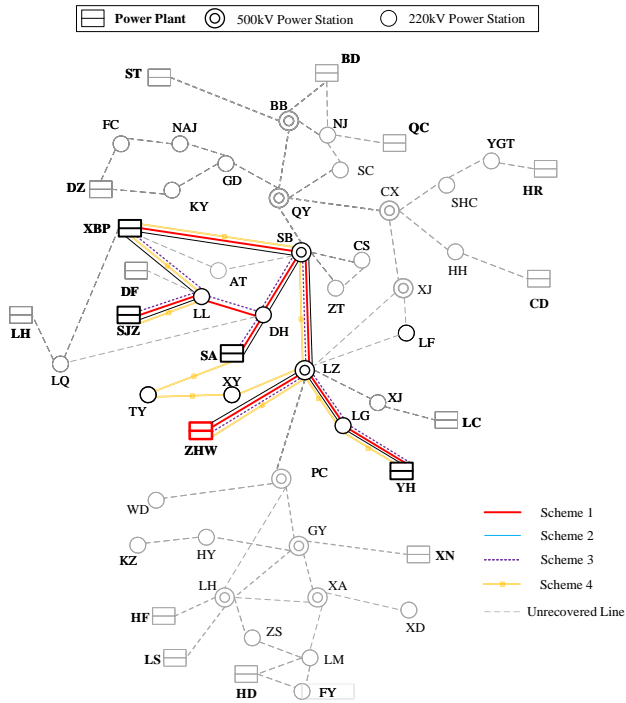

*Fig.9 Partial network of the southern Hebei power system and schematic diagram of alternative schemes*

## 6. Conlusions

During the system restoration process, the optimisation and decision making of energising path schemes is a key issue to realize unit and load restoration safely and efficiently. In general, the energising path scheme made beforehand may not be executed successfully due to the possible faults on the related lines under the extreme system condition, so it is necessary to provide alternative energising path schemes for the dispatchers. In view of this, a novel path optimisation method is presented in this paper according to the idea of determining the alternative path scheme set first and then sorting the schemes. The proposed iterative searching method for alternative path schemes could determine a group of schemes with minimal charging reactive power efficiently. And the evaluation method can sort the alternative schemes rationally and comprehensively. The New England 10-unit 39-bus system and the southern Hebei power system of China are employed to demonstrate the effectiveness of the proposed method. The results show that the proposed method can effectively determine a group of alternative path schemes with good calculation performance for power system restoration.

## 7. Acknowledgments

The work in this paper was supported by a project of National Natural Science Foundation of China (51677071), and a project of the Fundamental Research Funds for the Central Universities (2018MS085).

## 8. References

1. Ancona, J.J.: 'A Framework for Power System Restoration Following a Major Power Failure', *IEEE Trans. Power Syst.*, 1995, **10**, (3), pp. 1480-1485
2. Fink, L.H., Liou, K.L., Liu, C. C.: 'From Generic Restoration Actions to Specific Restoration Strategies', *IEEE Trans. Power Syst.*, 1995, **10**, (2), pp. 745-752
3. Sun, W., Liu, C.C.: 'Optimal Transmission Path Search in Power System Restoration', in, *Proc. 2013 IREP Symposium Bulk Power System Dynamics and Control - IX Optimization, Security and Control of the Emerging Power Grid*, Rethymno, Greece, August 2013, pp. 1-5
4. Adibi, M.M., Alexander, R.W., Milanicz, D.P.: 'Energizing High and Extra-High Voltage Lines During Restoration', *IEEE Trans. Power Syst.*, 1999, **14**, (3), pp. 1121-1126
5. Wang, C., Vittal, V., Kolluri, V.S., *et al.*: 'PTDF-Based Automatic Restoration Path Selection', *IEEE Trans. Power Syst.*, 2010, **25**, (3), pp. 1686-1695
6. Lin, Z., Wen, F., Xue, Y.: 'A Restorative Self-Healing Algorithm for Transmission Systems Based on Complex Network Theory', *IEEE Trans. Smart Grid*, 2017, **7**, (4), pp. 2154-2162
7. Chung, F.C.: 'Computer-Aided Energy System Restoration Automation', PhD thesis, The University of Texas at Arlington, 1986
8. Hou, Y., Liu, C.C., Sun, K., *et al.*: 'Computation of Milestones for Decision Support During System Restoration', *IEEE Trans. Power Syst.*, 2011, **26**, (3), pp. 1-10
9. Liu, Y., Sun, P., Wang, C.: 'Group Decision Support System for Backbone-Network Reconfiguration', *Int. J. Electr. Power Energy Syst.*, 2015, **71**, (6), pp. 391-402
10. Zhou, Y., Min, Y.: 'Optimal Algorithm for System Reconstruction', in, *Proc. Int. Conf. on Power System Technology*, Kunming, China, December 2002, pp. 201-203
11. Dimitrijevic, S., Rajakovic, N.: 'An Innovative Approach for Solving the Restoration Problem in Distribution Networks', *Electr. Power Syst. Res.*, 2011, **81**, (10), pp. 1961-1972
12. Gu, X., Zhong, H.: 'Optimisation of Network Reconfiguration Based on a Two-Layer Unit-Restarting Framework for Power System Restoration', *IET Gener. Transm. Distrib.*, 2012, **6**, (7), pp. 693-700
13. Cao, X., Wang, H., Liu, Y., *et al.*: 'Coordinating Self-Healing Control of Bulk Power Transmission System Based on a Hierarchical Top-Down Strategy', *Int. J. Electr. Power Energy Syst.*, 2017, **90**, pp. 147-157
14. Zhang, C., Lin, Z., Wen, F., *et al.*: 'Two-Stage Power Network Reconfiguration Strategy Considering Node Importance and Restored Generation Capacity', *IET Gener. Transm. Distrib.*, 2014, **8**, (1), pp. 91-103
15. Shi, L.B., Ding, H.L., Xu, Z.: 'Determination of Weight Coefficient for Power System Restoration', *IEEE Trans. Power Syst.*, 2012, **27**, (2), pp. 1140-1141
16. Yun, Z., Zheng, Y., Naihu, L., *et al.*: ' A New System Restoration Path Search Algorithm and Its Applicability Research ', *Proc. of the CSEE*, 2016, **36**, (15), pp. 4152-4161
17. Sun, W., Liu, C.-C., Zhang, L.: 'Optimal Generator Start-up Strategy for Bulk Power System Restoration', *IEEE Trans. Power Syst.*, 2011, **26**, (3), pp. 1357-1366
18. Sun, L., Zhang, C., Lin, Z., *et al.*: 'Network Partitioning Strategy for Parallel Power System Restoration', *IET Gener. Transm. Distrib.*, 2016, **10**, (8), pp. 1883-1892
19. Jiang, Y., Chen, S., Liu, C.C., *et al.*: 'Blackstart Capability Planning for Power System Restoration', *Int. J. Electr. Power Energy Syst.*, 2017, **86**, pp. 127-137
20. Sun, L., Lin, Z., Xu, Y., *et al.*: 'Optimal Skeleton-Network Restoration Considering Generator Start-up Sequence and Load Pickup', *IEEE Trans. Smart Grid*, 2018, PP, (99), pp. 1-1
21. Fisher, E.B., Neill, R.P., Ferris, M.C.: 'Optimal Transmission Switching', *IEEE Trans. Power Syst.*, 2008, **23**, (3), pp. 1346-1355
22. Song, K., Xie, Y., Yin, M., *et al.*: ' Mixed Integer Linear Optimization Model for Path Restoration of Blackout System Based on Network Flow Theory', *Autom. of Electr. Power Syst.*, 2017, **41**, (3), pp. 25-32
23. Xie, Y., Song, K., Wu, Q.: 'Orthogonal Genetic Algorithm Based Power System Restoration Path Optimization', *Int. Trans. Electr. Energy Syst.*, 2018

24. Bretas, A.S., Phadke, A.G.: 'Artificial Neural Networks in Power System Restoration', *IEEE Power Eng. Rev.*, 2002, **22**, (10), pp. 61-61
25. Ma, T.K., Liu, C.C., Tsai, M.S., *et al.*: 'Operational Experience and Maintenance of Online Expert System for Customer Restoration and Fault Testing', *IEEE Trans. Power Syst.*, 1992, **7**, (2), pp. 835-842
26. Adibi, M., Kafka, R., Milanicz, D.: 'Expert System Requirements for Power System Restoration', *IEEE Trans. Power Syst.*, 1994, **9**, (3), pp. 1592-1600
27. Teo, C.Y., Wei, S.: 'Development of an Interactive Rule-Based System for Bulk Power System Restoration', *IEEE Trans. Power Syst.*, 2000, **15**, (2), pp. 646-653
28. Prömel, H.J., Steger, A.: 'The Steiner Tree Problem : A Tour through Graphs, Algorithms, and Complexity' (Friedr. Vieweg & Sohn Press, 2002)
29. Liu, Q., Shi, L., Ni, Y., *et al.*: 'Intelligent Optimization Strategy of the Power Grid Reconfiguration During Power System Restoration', *Proc. of the CSEE*, 2009, **29**, (13), pp. 8-15
30. Wang D., Gu X., Zhou G., *et al.*: 'Decision-making optimization of power system extended black-start coordinating unit restoration with load restoration'. *Int. Trans. Electr. Energy Syst.*, 2017, **27,** (1), pp. 1-18
31. Ortega, F., Wolsey, L.A.: 'A Branch-and-Cut Algorithm for the Single-Commodity, Uncapacitated, Fixed-Charge Network Flow Problem †', *Networks*, 2003, **41**, (3), pp. 143–158
32. Floudas, C.A.: *Nonlinear and Mixed-Integer Optimization: Fundamentals and Applications*, (Oxford University Press, 1995)
33. Liu, W., Lin, Z., Wen, F., *et al.*: 'Analysis and Optimisation of the Preferences of Decision-Makers in Black-Start Group Decision-Making', *IET Gener. Transm. Distrib.*, 2013, **7**, (1), pp. 14-23
34. Liu, Y., Fan, R., Terzija, V.: 'Power System Restoration: A Literature Review From 2006 to 2016', *J. Mod. Power Syst. Clean Energy*, 2016, **4**, (3), pp. 332-341
35. Feltes, J.W., Grande-Moran, C., Duggan, P., *et al.*: 'Some Considerations in the Development of Restoration Plans for Electric Utilities Serving Large Metropolitan Areas', *IEEE Trans. Power Syst.*, 2006, **21**, (2), pp. 909-915
36. Liu, Y., Gu, X.: 'Skeleton-Network Reconfiguration Based on Topological Characteristics of Scale-Free Networks and Discrete Particle Swarm Optimization', *IEEE Trans. Power Syst.*, 2007, **22**, (3), pp. 1267-1274
37. Ma, J., Fan, Z.P., Huang, L.H.: 'A Subjective and Objective Integrated Approach to Determine Attribute Weights', *Eur. J. Oper. Res.*, 1999, **112**, (2), pp. 397-404
38. Ke, H., Liu, S.: 'Similarity to Ideal-Based Grey Relational Projection for Multiple Criteria Decision-Making', in, *Proc. IEEE Int. Conf. on Grey Systems and Intelligent Services*, Nanjing, China, November 2009, pp.1008-1012
39. Rosenthal, R. E.: 'GAMS - a user's guide' (GAMS Development Corporation, 2010)
40. Gu, X., Liu, W., Sun, C.: ' Optimisation for unit restarting sequence considering decreasing trend of unit start-up efficiency after a power system blackout ', *IET Gener. Transm. Distrib.*, 2016, **10**, (16), pp. 4187-4196